\documentclass[conference]{IEEEtran}
\IEEEoverridecommandlockouts
\usepackage{cite}
\usepackage{amsmath,amssymb,amsfonts}
\usepackage{algorithmic}
\usepackage{graphicx}
\usepackage{textcomp}
\usepackage{xcolor}
\usepackage{tabularx}
\usepackage{tikz}
\usepackage{url}

\usepackage{cuted}
\usetikzlibrary{arrows.meta,positioning,calc}
\usepackage{amsthm}
\usepackage{amsmath,amsthm,amsfonts,amssymb,amscd}
\theoremstyle{definition}

\def\BibTeX{{\rm B\kern-.05em{\sc i\kern-.025em b}\kern-.08em
    T\kern-.1667em\lower.7ex\hbox{E}\kern-.125emX}}

\usepackage{algorithm,algorithmic}

\usepackage{fancyhdr}

\begin{document}

\title{\LARGE \textbf{PIVONet: A Physically-Informed Variational Neuro ODE Model for Efficient Advection-Diffusion Fluid Simulation} }

\author{\IEEEauthorblockN{Hei Shing Cheung}
\IEEEauthorblockA{\textit{Division of Engineering Science} \\
\textit{University of Toronto}\\
Toronto, Canada \\
hayson.cheung@mail.utoronto.ca}
\and
\IEEEauthorblockN{Qicheng Long}
\IEEEauthorblockA{\textit{Division of Engineering Science} \\
\textit{University of Toronto}\\
Toronto, Canada \\
ethan.long@mail.utoronto.ca}
\and
\IEEEauthorblockN{Zhiyue Lin}
\IEEEauthorblockA{\textit{Division of Engineering Science} \\
\textit{University of Toronto}\\
Toronto, Canada \\
davidlzy.lin@mail.utoronto.ca}
}

\maketitle

\begin{abstract}
   We present PIVONet (\emph{P}hysically-\emph{I}nformed \emph{V}ariational \emph{O}DE \emph{N}eural \emph{Net}work), a unified framework that integrates Neural Ordinary Differential Equations (Neuro-ODEs) with Continuous Normalizing Flows (CNFs) for stochastic fluid simulation and visualization. 
   First, we demonstrate that a physically informed model, parameterized by CNF parameters $\theta$, can be trained offline to yield an efficient surrogate simulator for a specific fluid system, eliminating the need to simulate the full dynamics explicitly.
   Second, by introducing a variational model with parameters $\phi$ that captures latent stochasticity in observed fluid trajectories, we model the network output as a variational distribution and optimize a pathwise Evidence Lower Bound (ELBO), enabling stochastic ODE integration that captures turbulence and random fluctuations in fluid motion (advection-diffusion behaviors).

\end{abstract}

\begin{figure}[h!]
  \centering

  \begin{minipage}[b]{0.3\linewidth}
    \centering
    \includegraphics[width=\linewidth]{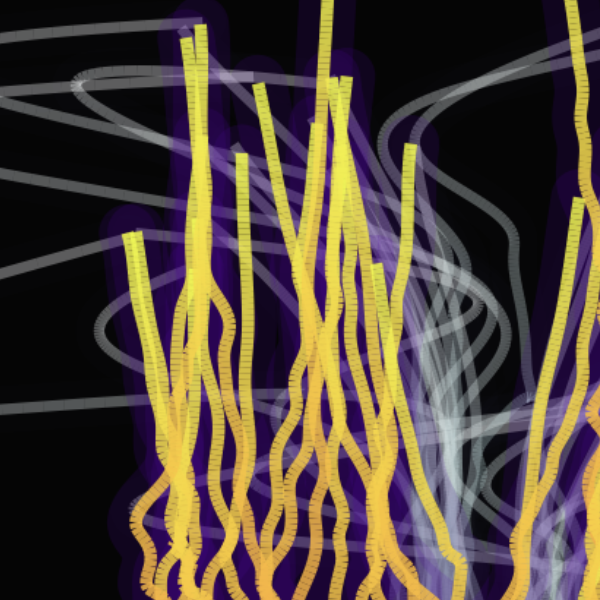}
    \vspace{2pt}
    {\small (a)}
  \end{minipage}
  \hspace{0.015\linewidth}
  \begin{minipage}[b]{0.3\linewidth}
    \centering
    \includegraphics[width=\linewidth]{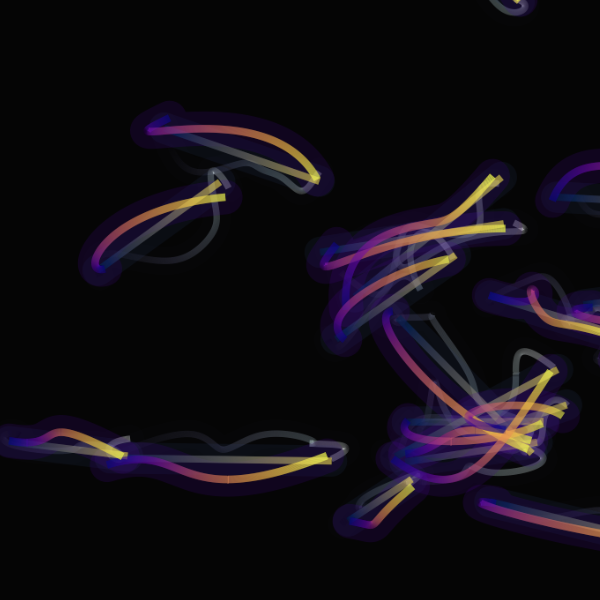}
    \vspace{2pt}
    {\small (b)}
  \end{minipage}
  \hspace{0.015\linewidth}
  \begin{minipage}[b]{0.3\linewidth}
    \centering
    \includegraphics[width=\linewidth]{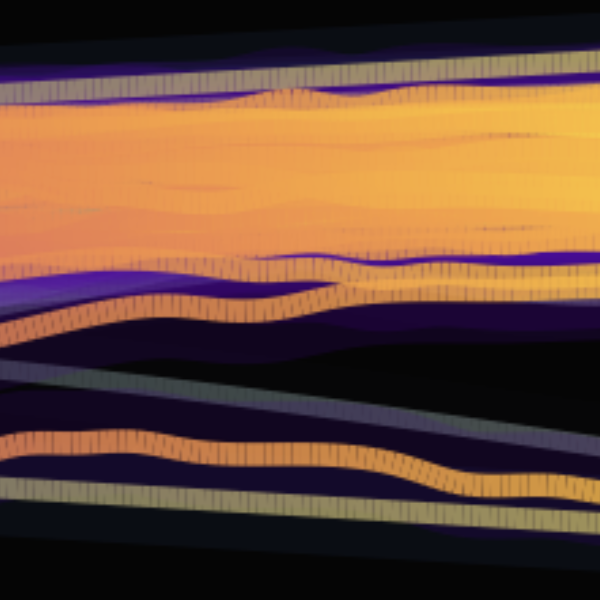}
    \vspace{2pt}
    {\small (c)}
  \end{minipage}

  \caption{Generated flow trajectories overlaid onto ground truth:  
  (a) target-matching under vortex shearing,
  (b) sparse particle regions, and  
  (c) flow separation.}
  \label{fig:three-side-by-side}
\end{figure}

\section{Introduction}

The motion of particles in fluids is governed by complex interactions between the underlying flow field and stochastic perturbations arising from thermal effects or turbulence~\cite{Squires2005}. Traditional computational fluid dynamics (CFD) methods provide high-fidelity simulations but are often prohibitively expensive for real-time prediction or large-scale parameter studies. 

Whereas our data are generated from imperfect computational simulations of particle trajectories governed by fluid mechanical equations, we use this as an opportunity to explore whether latent dynamical models can capture key flow features in non-laminar regimes, including advection--diffusion, turbulence, and flow separation. The goal is to assess whether such models can efficiently predict particle trajectories in two-dimensional fluid flows. We present a computationally tractable alternative for applications where traditional CFD methods are impractical.

While our model is trained on a limited set of flow conditions, it demonstrates that latent representations can capture the key patterns present within this regime. Although we do not claim broad generalization, this provides a foundation on which broader, covariate-conditioned generalization could be built in future work.

\section{Background}
\subsection{Governing Advection-Diffusion  Equation}
\begin{figure}[h!]
    \centering
    \includegraphics[width=1\linewidth]{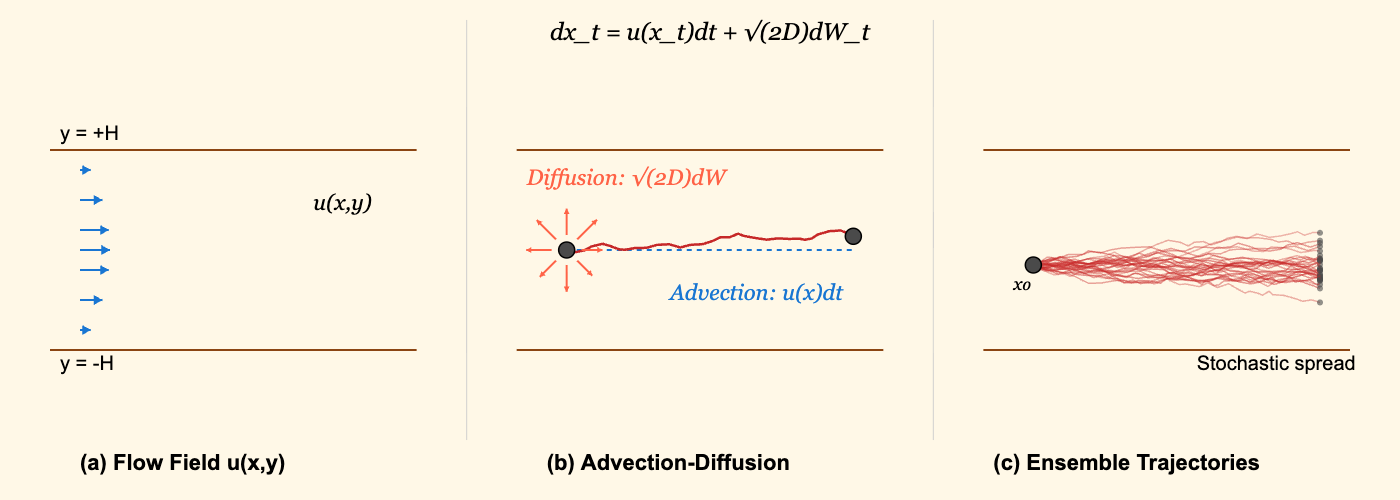}
    \caption{\textbf{Langevin Dynamics: Particle Motion in Fluid Flow.} The Langevin equation
    $dx_t = u(x_t)dt + \sqrt{2D}dW_t$ governs particle motion in fluid flows.
(a) Poiseuille velocity profile.
(b) Single particle trajectory.
(c) Multiple stochastic realizations from the same initial condition.
}
    \label{fig:cfd_langevin}
\end{figure}

We model particle trajectories using the overdamped Langevin equation, which captures advection-diffusion physics while remaining computationally tractable~\cite{Ermak1978}. The position $\mathbf{x}_t = (x_t, y_t) \in \mathbb{R}^2$ evolves according to:
\begin{equation}
    d\mathbf{x}_t = \mathbf{u}(\mathbf{x}_t) \, dt + \sqrt{2D} \, d\mathbf{W}_t
    \label{eq:langevin}
\end{equation}
This is a SDE where $\mathbf{u}(\mathbf{x}_t)$\footnote{The simulation to obtain $u$ will be discussed in Section \ref{sec:xsimulation}} is the local fluid velocity field (Fig.~\ref{fig:cfd_langevin}a), $D$ is the diffusion coefficient, and $\mathbf{W}_t$ is a two-dimensional Brownian motion, neglecting inertial effects. 

 The SDE in Eq.~\eqref{eq:langevin} decomposes particle motion into deterministic advection along streamlines and stochastic diffusion (Fig.~\ref{fig:cfd_langevin}b). The equation is discretized by Euler-Maruyama:
\begin{equation}
    \mathbf{x}_{t+\Delta t} = \mathbf{x}_t + \mathbf{u}(\mathbf{x}_t) \, \Delta t + \sqrt{2D \, \Delta t} \, \boldsymbol{\xi}, \quad \boldsymbol{\xi} \sim \mathcal{N}(0, \mathbf{I})
    \label{eq:euler_maruyama}
\end{equation}
with reflecting boundaries at $y = \pm H$. It is mathematically analogous to lattice diffusion systems expressed as linear ODEs of the form 
\begin{equation}
    \mathbf{x}'(t) = k A \mathbf{x}(t)
\end{equation}
with reflecting endpoints, whose eigenvalue structure governs long-term equilibrium behavior~\cite{ODETextbook}.
 
The stochastic nature of this process leads to variability in particle trajectories even from identical initial conditions (Fig.~\ref{fig:cfd_langevin}c), consistent with classical dispersion theory~\cite{Taylor1953}. For each trajectory, we record initial conditions $\mathbf{x}_0$, flow parameters $\mathbf{\alpha}$, and the sequence $\{\mathbf{x}_t\}_{t=0}^{T}$.

The flow profiles used in this paper are displayed in Appendix~\ref{apx:figures} under Figure~\ref{fig:flow_vis}.

\subsection{Normalizing Flow}
For trajectory modeling, continuous normalizing flows (CNFs) treat the transformation as an ODE~\cite{Chen2018,Grathwohl2019}. This builds upon earlier discrete flow-based generative models such as NICE~\cite{Dinh2014}, RealNVP~\cite{Dinh2017}, and Glow~\cite{Kingma2018}, originally introduced for density estimation and variational inference~\cite{Rezende2015}.
\begin{equation}
    \frac{d\mathbf{z}(t)}{dt} = g_{\boldsymbol{\theta}}(\mathbf{z}(t), t), \quad \mathbf{z}(0) = \mathbf{z}_0, \quad \mathbf{z}(1) = \mathbf{x}
    \label{eq:cnf_ode}
\end{equation}
The log-density evolves according to the instantaneous change-of-variables formula~\cite{Chen2018}:
\begin{equation}
    \frac{d \log p(\mathbf{z}(t))}{dt} = -\text{tr}\left(\frac{\partial g_{\boldsymbol{\theta}}}{\partial \mathbf{z}}\right)
    \label{eq:cnf_logprob_ode}
\end{equation}
This equation tracks how probability mass compresses or expands as particles flow through the learned transformation. The trace of the Jacobian corresponds to the divergence of the velocity field and governs local density expansion or contraction. For our particle trajectories, this enables the model to learn where paths concentrate (high-probability regions like flow streamlines) versus where they disperse (low-probability regions affected by strong diffusion)~\cite{Grathwohl2019}.
\subsection{Bi-directionality and Training}
\begin{figure}[h!]
    \centering
    \includegraphics[width=1\linewidth]{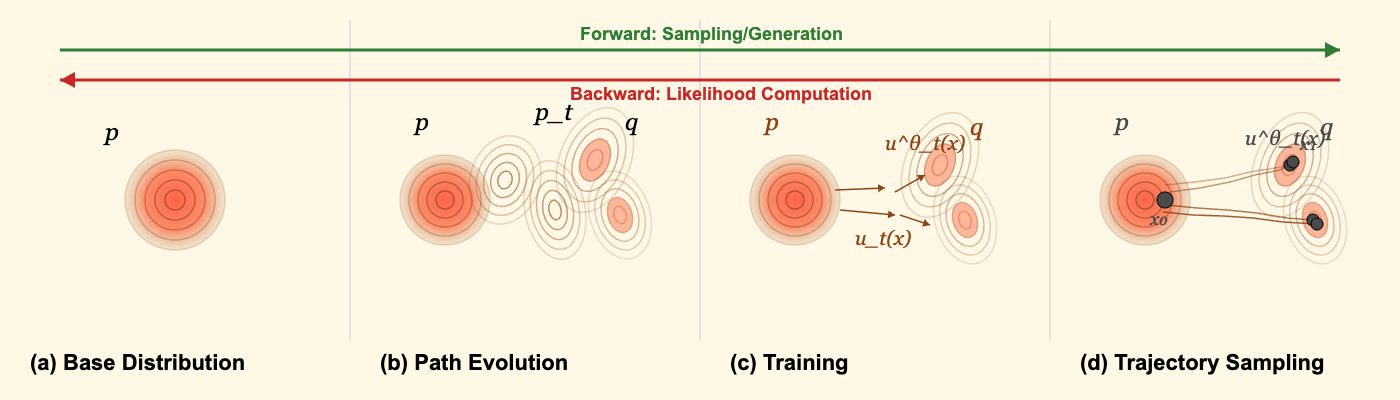}
    \caption{\textbf{Continuous Normalizing Flow for Particle Trajectory Prediction
.} The CNF transforms a simple Gaussian base distribution $p$ (a) through intermediate distributions (b) via learned velocity fields $u_t(x)$ (c) to match the complex trajectory distribution $q$. During sampling (d), particles follow stochastic paths from initial position $x_0$ to predicted position $x_1$, capturing both deterministic flow and diffusive behavior.}
    \label{fig:cnf_transformation}
\end{figure}

Since we model the mappings between the two distributions as a diffeomorphism, we have an advantage of CNF being bi-directionality: we can integrate the ODE system both forward (from simple $\mathbf{z}_0 \sim \mathcal{N}(0, \mathbf{I})$ to complex trajectory distribution) and backward (from observed trajectory to its latent representation). The forward process generates new trajectories by sampling, while the backward process computes exact likelihoods for training~\cite{Rezende2015}. Figure~\ref{fig:cnf_transformation} illustrates this bidirectional transformation, showing how the initially Gaussian distribution evolves through intermediate time steps into the complex, multimodal distribution of particle trajectories conditioned on flow parameters. The ODE system in Eqs.~\eqref{eq:cnf_ode}--\eqref{eq:cnf_logprob_ode} is solved jointly using adaptive solvers, allowing us to compute exact likelihoods for any trajectory under the learned distribution—essential for training and uncertainty quantification~\cite{Kingma2018}.

We condition on $\mathbf{x}_0$ and flow parameters $\{U_0, \gamma, U{\text{max}}, H, D, \ldots \} \subset \boldsymbol{\alpha}$ via concatenation or feature-wise transformations. The model is trained by maximizing log-likelihood:
\begin{equation}
    \mathcal{L}(\boldsymbol{\theta}) = \mathbb{E}_{\mathbf{x} \sim p_{\text{data}}} \left[ \log p_{\boldsymbol{\theta}}(\mathbf{x} \mid \mathbf{x}_0, \boldsymbol{\alpha}) \right]
    \label{eq:nf_objective}
\end{equation}

For inference, we sample $\mathbf{z} \sim \mathcal{N}(0, \mathbf{I})$ and apply $\hat{\mathbf{x}} = f_{\boldsymbol{\theta}}(\mathbf{z}; \mathbf{x}_0, \boldsymbol{\alpha})$ to obtain predicted trajectories without re-running simulations. This framework captures both deterministic flow-driven behavior and stochastic variability, enabling probabilistic trajectory forecasting.

\section{Model}
\begin{figure*}[h!]
    \centering
    \includegraphics[width=1\linewidth]{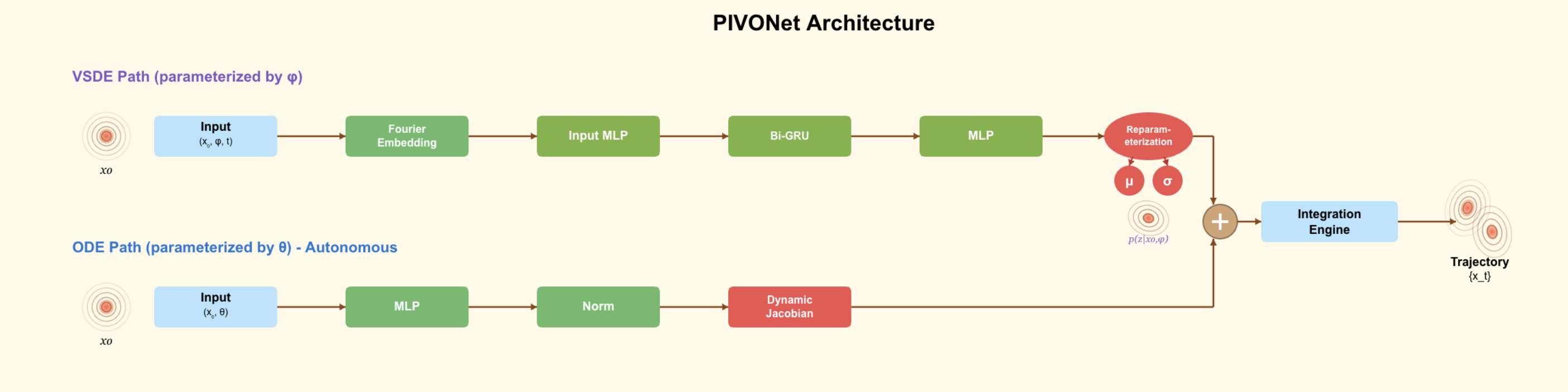}
    \caption{The VSDE framework couples a frozen CNF backbone with learned posterior controls and physics-informed losses.}
    \label{fig:architecture}
\end{figure*}

\subsection{Neural Differential Equation Backbone}
We train a continuous normalizing flow (CNF) to inherit the backbone dynamics of the flow \cite{Chen2018,Grathwohl2019}. The CNF delivers an expressive drift \(f_\theta(z)\) that encodes the physics of the training dataset, while the VSDE extends it with stochastic diffusion and controls to match observed trajectories. Afterwards, we treat the CNF as a frozen module and learn only the additional control signal \(u(z, t)\) and diffusion coefficient \(g\), which keeps the base physics intact while providing flexibility during inference.
Adding the VSDE path is therefore equivalent to augmenting the modeling ODE with a nonhomogeneous stochastic forcing term, letting the learned controls and diffusion absorb dataset-specific deviations without corrupting the pretrained drift.
\begin{equation}
\mathrm{d}z = f_\theta \,\mathrm{d}t + g_\phi\,\mathrm{d}W_t 
\end{equation}
Here the control \(u(z, t)\) and diffusion coefficient \(g\) act as the learned forcing, while \(\mathrm{d}W_t\) denotes the Brownian increment that encodes stochasticity.

By adding the variational component, our model tends to explore different pathways when reconstructing the same input; This results in a higher stability as the model is more likely to stay on the learned manifold, as the variational component can help nudge the trajectory back when it starts to drift away.

\subsection{Architecture}
Input trajectories are first encoded by the Trajectory Encoder, which produces a compact posterior context vector. This context initializes a surrogate posterior \(q(z_0\mid x)\) and conditions the Posterior Drift Net. The posterior drift network, implemented as an MLP with Fourier time embeddings, outputs both the control vector \(u\) and optional diffusion corrections \cite{Goodfellow2016}. The VSDE integrator then combines the fixed CNF drift, the learned control, and the Brownian diffusion to produce latent trajectories that are mapped back to the original space for reconstruction and diagnostics (see Figure~\ref{fig:architecture}).

\subsection{Activation Function}
The VSDE control MLP uses Gaussian Error Linear Units (GELU) in each hidden layer, which blend linear and nonlinear behaviors and keep gradients smooth so the control signal changes gradually with the latent state. The GRU encoder relies on sigmoid activations for the reset and update gates and a tanh nonlinearity for the candidate hidden state; this combination ensures gate outputs stay in \([0,1]\) and that the recurrent dynamics remain bounded while still allowing nonlinear transformations of the inputs, along with why these activations are preferred for time-series modeling and function approximation.

\subsection{Variational Inference}
We perform variational inference by sampling \(n_{\text{particles}}\) latent initializations from \(q(z_0\mid x)\) and integrating each sample through the VSDE \cite{Rezende2015}. The ELBO decomposes into a reconstruction likelihood (driven by observation noise \(\sigma_{\text{obs}}\)), a KL divergence that anchors the posterior to a standard normal, and a control cost that penalizes large control efforts. The energy-based penalties described below are added to this objective to shepherd the latent trajectories toward physically consistent behavior.
Sampling uses the reparameterization trick:
\begin{equation}
z_0 = \mu + \sigma \odot \epsilon,\quad \epsilon \sim \mathcal{N}(0, I),
\end{equation}
which lets gradients flow through the encoder parameters \((\mu, \sigma)\) when we backpropagate the loss to optimize the neural network. The gradient of the ELBO with respect to the parameters \(\theta\) is
\begin{align}
\nabla_\theta \mathcal{L} = \mathbb{E}_{q(z_0\mid x)}\left[\nabla_\theta \log p(x\mid z) - \lambda_{\text{KL}} \nabla_\theta \log q(z_0\mid x) + \cdots\right],
\end{align}
where the ellipsis hides control and physics loss derivatives and the expectation is approximated with our particles. The adjoint method ensures the gradients through the VSDE integrator remain tractable.

\subsection{Training}
\subsubsection{Optimizing the Path}
We backpropagate through the integrator using the adjoint sensitivity method built into the CNF. The VSDE integrator is differentiable because we keep the time grid fixed and apply deterministic updates before adding the diffusion term. The optimizer updates the Posterior Drift Net weights, the diffusion log-parameter, and the posterior encoder simultaneously.

\subsubsection{Dynamic Jacobian}
Because the CNF defines an instantaneous velocity field, we can compute the Jacobian of the combined drift with respect to \(z\) during training. This Jacobian is needed for KL and control cost terms as well as for ensuring stability when coupling the CNF with the posterior drift.

\subsubsection{Physical Loss Function}
We augment the ELBO with two complementary physics-aware losses so the VSDE latent paths respect conserved quantities and known PDE structure. The first loss tracks kinetic energy fluctuations by computing \(\frac{1}{2}\rho\|z_{t+1} - z_t\|^2\) along each trajectory and penalizing sudden jumps that contradict the smooth advection-dominated physics; this couples adjacent timesteps through finite differences of latent velocities. The second loss enforces the residual of the compressible energy equation
\(\rho c_p(\partial_t T + u \cdot \nabla T) = k \nabla^2 T + \Phi\)
by discretizing the temporal derivatives via backward differences and estimating spatial gradients from the CNF decoder; penalizing this residual steers the diffusion and control terms to satisfy thermodynamic balance in expectation. Each term is averaged over particles before being added to the loss, and we scale them with \(\lambda_{\text{phys}}\) and \(\lambda_{\text{pde}}\) so the model can trade reconstruction fidelity for these physical regularizers.

\subsubsection{Joint Evidence Lower-Bound}
The final loss is
\begin{multline}
\mathcal{L} = -\mathbb{E}_{q(z_0\mid x)}\left[ \log p(x\mid z) \right] + \lambda_{\text{KL}} \mathrm{KL}(q(z_0\mid x) \| p(z_0)) \\ + \lambda_{u} \mathcal{L}_{\text{control}} + \lambda_{\text{phys}} \mathcal{L}_{\text{energy}} + \lambda_{\text{pde}} \mathcal{L}_{\text{balance}}.
\end{multline}
The \(\lambda\) weights for the KL, control, and physics penalties are instead selected manually in this study to balance numerical stability with expressive control rather than through an exhaustive grid search.

\subsubsection{Hyperparameters and Optimizers}
Training uses the Adam optimizer as described in \cite{adam}, with weight decay (AdamW) and cosine learning-rate annealing. The learning rate schedule is warmed up for the first 2k steps and decays toward 1e-5. Control cost scale and diffusion coefficients are initialized to encourage low-variance trajectories, and the physics weights start near zero to prevent destabilizing gradients.

\subsection{Inference}
\subsubsection{Joint ODE Integration}
Inference samples from the encoder to obtain posterior contexts and then integrates the VSDE for both control and diffusion to produce latent trajectories. The output is decoded via the CNF to reconstruct the observable variables while preserving uncertainty quantification through the ensemble of particles.

\subsubsection{Integration Methods}
We expose Euler, Heun, RK4, and Dormand–Prince integrators for the deterministic part of the dynamics. Diffusion is always attached via Euler–Maruyama \eqref{eq:euler_maruyama}. The inference API allows selecting the integrator and diffusion scale so experiments can trade accuracy for speed, though emprical results show little differences in accuracy across integrators for our application.

\subsection{Guard Rails During Inference}
Special-case logic clamps controls and enforces absorbing boundaries when trajectories approach invalid states, which prevents numerical explosions. If a trajectory wanders outside the training manifold, its weight in the ELBO computation is attenuated to focus the optimization on valid samples.
We formalize this by defining a soft validity indicator
\begin{equation}
\gamma(z) = \exp\left(-\alpha \max\left(\|z\| - R, 0\right)^2\right),
\end{equation}
where \(R\) is a radius extracted from the training states and \(\alpha\) controls how sharply samples are down-weighted. The encoder weight is multiplied by \(\gamma(z)\) before being averaged, and the posterior control is clamped according to
\begin{equation}
u_{\text{clamp}} = \text{sign}(u) \min\left(|u|, u_{\text{max}}\right),
\end{equation}
ensuring the VSDE cannot inject unbounded drifts even when extrapolating.

\subsection{Algorithmic Description}
\label{sec:algorithm}
Algorithm~\ref{alg:pivonet} summarizes the complete training and inference procedure for PIVONet. The method first pretrains a continuous normalizing flow (CNF) backbone, then augments it with a Variational Stochastic Differential Equation (VSDE) controller to capture stochastic dynamics.

\begin{algorithm}[h!]
\caption{PIVONet: Neural Network Training}
\label{alg:pivonet}
\begin{algorithmic}[1]
\REQUIRE Training trajectories $\{x^{(i)}\}_{i=1}^{N}$, flow parameters $\alpha$, time grid $\{t_k\}_{k=0}^{T}$, learning rate $\eta$

\STATE \textbf{Stage 1: CNF Pretraining}
\STATE Initialize $\theta$ randomly
\REPEAT
\STATE Sample minibatch $\mathcal{B}$; sample $z^{(i)} \sim \mathcal{N}(0, I)$
\STATE $\hat{x}^{(i)} \gets f_\theta(z^{(i)}, \alpha)$; compute $\mathcal{L}_{\mathrm{CNF}} = \frac{1}{B}\sum_i \|x^{(i)} - \hat{x}^{(i)}\|_2^2$
\STATE Update $\theta$ via Adam
\UNTIL{CNF converged}
\STATE Freeze $\theta$ to obtain $f_\theta(\cdot)$

\STATE \textbf{Stage 2: VSDE Training}
\STATE Initialize $\psi, \phi, g$ randomly
\REPEAT
\STATE Sample minibatch $\mathcal{B}$; compute $(\mu, \sigma) \gets \mathrm{Encoder}_\psi(\mathcal{B})$
\FOR{$j = 1$ \TO $n_p$}
\STATE $z_0^{(j)} \gets \mu + \sigma \odot \mathcal{N}(0, I)$
\FOR{$k = 0$ \TO $T-1$}
\STATE $(u_k^{(j)}, \tilde{g}_k^{(j)}) \gets \mathrm{DriftNet}_\phi(z_k^{(j)}, t_k, (\mu, \sigma))$
\STATE $z_{k+1}^{(j)} \gets z_k^{(j)} + f_\theta(z_k^{(j)}) \Delta t + u_k^{(j)} \Delta t + \tilde{g}_k^{(j)} \Delta W_k$
\ENDFOR
\ENDFOR
\STATE Compute: $\mathcal{L}_{\mathrm{recon}}, \mathcal{L}_{\mathrm{KL}}, \mathcal{L}_{\mathrm{control}}, \mathcal{L}_{\mathrm{energy}}, \mathcal{L}_{\mathrm{pde}}$
\STATE $\mathcal{L} = \mathcal{L}_{\mathrm{recon}} + \lambda_{\mathrm{KL}} \mathcal{L}_{\mathrm{KL}} + \lambda_u \mathcal{L}_{\mathrm{control}} + \lambda_{\mathrm{phys}} \mathcal{L}_{\mathrm{energy}} + \lambda_{\mathrm{pde}} \mathcal{L}_{\mathrm{pde}}$
\STATE Update $(\psi, \phi, g)$ via Adam
\UNTIL{ELBO converged}
\end{algorithmic}
\end{algorithm}
\section{Experiment Methodology}
\subsection{Hypothesis}
The experiment is designed to test the hypothesis that:
\begin{enumerate}
    \item Using neuro ODE-based generative models can effectively capture complex fluid dynamics.
    \item Integrating VSDE controllers with CNF backbones enhances the model's ability to represent stochastic behaviors in fluid flows.
    \item The proposed framework can generalize across different flow regimes, including inviscid, viscous, and incompressible scenarios.
\end{enumerate}

\subsection{Data Simulation}\label{sec:xsimulation}

To generate high-fidelity, time-resolved flow fields for training and evaluation of continuous normalizing flow (CNF) and stochastic differential equation (SDE) models, we employed the open-source \texttt{PyFR} framework as the numerical simulation engine~\cite{witherden2025pyfr}. \texttt{PyFR} is a Python-based computational fluid dynamics (CFD) platform that implements high-order accurate spatial discretization using the flux reconstruction approach and is capable of solving the compressible Navier–Stokes equations on unstructured meshes across heterogeneous hardware architectures.

The specifications for the \texttt{PyFR} orchestrations is in Appendix \ref{apx:pyfr}

The governing conservation laws for compressible fluid flow are expressed as:

\begin{equation}\label{eq:Navier-Stokes}
\frac{\partial \mathbf{U}}{\partial t}
+ \nabla \cdot \mathbf{F}(\mathbf{U})
= \nabla \cdot \mathbf{F}_v(\mathbf{U}, \nabla \mathbf{U}),
\end{equation}

where \(\mathbf{U}\) denotes the vector of conserved variables, \(\mathbf{F}\) and \(\mathbf{F}_v\) represent inviscid and viscous fluxes respectively, and the right-hand side encapsulates viscous transport contributions.

Spatial discretization in \texttt{PyFR} employs a high-order flux reconstruction scheme that replaces spatial derivatives with discrete algebraic operators evaluated at solution and flux points defined on the computational mesh. Following discretization of the spatial operators, the partial differential equation (PDE) system is reduced to a large system of ordinary differential equations (ODEs):

\begin{equation}\label{eq:Simplified}
\frac{d\mathbf{u}(t)}{dt} = \mathbf{f}(\mathbf{u}(t)),
\end{equation}

where \(\mathbf{u}(t)\) is the vector of all discrete flow variables comprising density, momentum, and energy at each node, and \(\mathbf{f}\) encapsulates the semi-discrete flux and source approximations. \texttt{PyFR} advances this system in time using explicit time integration schemes suitable for high-order methods.

Because the resulting solutions still encode conserved quantities, they become the reference for our physics-informed loss. We later compute a transformed temperature field and velocity signal from the discrete states and penalize the residual of the thermal energy equation
\begin{equation}
\rho c_p\left(\frac{\partial T}{\partial t} + \mathbf{u} \cdot \nabla T\right) = k \nabla^2 T + \Phi,
\end{equation}
which balances temporal and advective heating against conduction and viscous dissipation. This augments the variational autoencoder with a PDE-based constraint that is consistent with the compressible simulation data.

\subsection{Flow Cases}
We evaluate our framework across three canonical fluid dynamics scenarios that exhibit diverse flow behaviors and complexities:
\begin{itemize}
    \item \textbf{2D Euler Vortex:} This case involves simulating a two-dimensional vortex in an inviscid fluid governed by the Euler equations. The vortex dynamics test the model's ability to capture rotational flow features and conserve vorticity over time.
    \item \textbf{Viscous Shock Tube:} This scenario simulates the evolution of shock waves and contact discontinuities in a viscous medium. The shock tube problem challenges the model to accurately represent sharp gradients and dissipative effects inherent in viscous flows.
    \item \textbf{Incompressible Cylinder Flow:} This case examines the flow around a circular cylinder in an incompressible fluid. The cylinder flow problem assesses the model's capability to capture boundary layer development, vortex shedding, and wake dynamics.
\end{itemize}

The specific configurations for each flow case, including domain size, mesh resolution, initial and boundary conditions, and physical parameters (e.g., Reynolds number, Mach number), are detailed in the code repository accompanying this manuscript. In addtional, below is a brief overview of the governing equations and numerical setups for each scenario.

\paragraph{2D Euler Vortex}
The 2D Euler vortex is governed by the inviscid Euler equations, which describe the conservation of mass, momentum, and energy in a compressible fluid without viscosity. The governing equations are expressed in conservation form as:
\begin{equation}
\frac{\partial \mathbf{U}}{\partial t} + \nabla \cdot \mathbf{F}(\mathbf{U}) = 0,
\end{equation}
where \(\mathbf{U}\) is the vector of conserved variables, and \(\mathbf{F}(\mathbf{U})\) represents the flux tensor. The initial condition consists of a Gaussian vortex superimposed on a uniform flow field. The computational domain is discretized using a structured mesh with periodic boundary conditions.

\paragraph{Viscous Shock Tube}
The viscous shock tube problem is governed by the compressible Navier–Stokes equations, which account for viscous effects in addition to the conservation of mass, momentum, and energy. The governing equations are given by:
\begin{equation}
\frac{\partial \mathbf{U}}{\partial t} + \nabla \cdot \mathbf{F}(\mathbf{U}) = \nabla \cdot \mathbf{F}_v(\mathbf{U}, \nabla \mathbf{U}),
\end{equation}
where \(\mathbf{F}_v(\mathbf{U}, \nabla \mathbf{U})\) represents the viscous flux tensor. The initial condition consists of a high-pressure region adjacent to a low-pressure region, separated by a diaphragm. The computational domain is discretized using an unstructured mesh with reflective boundary conditions at the tube walls.

\paragraph{Incompressible Cylinder Flow}
The incompressible cylinder flow is governed by the incompressible Navier–Stokes equations, which describe the conservation of mass and momentum in an incompressible fluid. The governing equations are expressed as:
\begin{subequations}
\begin{equation}
\nabla \cdot \mathbf{u} = 0,
\end{equation}
\begin{equation}
\frac{\partial \mathbf{u}}{\partial t} + \mathbf{u} \cdot \nabla \mathbf{u} = -\nabla p + \nu \nabla^2 \mathbf{u},
\end{equation}
\end{subequations}
where \(\mathbf{u}\) is the velocity field, \(p\) is the pressure field, and \(\nu\) is the kinematic viscosity. The initial condition consists of a uniform flow field impinging on a circular cylinder. The computational domain is discretized using an unstructured mesh with no-slip boundary conditions on the cylinder surface and far-field conditions at the domain boundaries.

\subsection{Experiment Workflow}
The empirical pipeline described below is executed across several flow cases (2D Euler vortex, viscous shock tube, and incompressible cylinder)  whose configurations are captured in the code base. Appendix \ref{apx:implementation} details the software implementation, which includes clearly defined workflow modules that matches the steps below. Training hyperparameters are provided in \ref{apx:hyperparams}

\begin{enumerate}
	\item \textbf{Trajectory simulation:} Particles are marched for 240 steps with varying \(\Delta t\) to produce bundles of trajectories, velocity snapshots, and mesh coordinates that subsequent modules reuse.
	\item \textbf{CNF backbone training:} The CNF drift network trains on the trajectory endpoints.
	\item \textbf{VSDE controller training:} With the CNF frozen, the variational posterior drift net optimizes the trajectories.
	\item \textbf{Inference diagnostics:} Ensembles of 64 particles over 120 steps are decoded and plotted alongside evaluation metrics.
\end{enumerate}

\subsection{Training and Experimental Metrics}\label{sec:training_metrics}
Data was simulated in bundles on the Google Cloud platfrom via an 80 GB NVIDIA H100 GPU cluster provided by Google Colab. Training runs used Apple M-Series silicon (MPS backend) so the CNF and VSDE modules finish within roughly 15 minutes each on the cached trajectory bundles; we warm up the learning rate for 1k steps and rely on AdamW with cosine annealing to stabilize convergence. The CFD trajectories are split into training and validation bundles that emphasize the vortex phase so the learned dynamics focus on the advection-dominated regime.

Hyperparameter values are summarized in Table~\ref{tab:hyperparams_euler}, \ref{tab:hyperparams_vis}, and \ref{tab:hyperparams_inc}  (Appendix~\ref{apx:hyperparams}). 
\subsection{Evaluation Sets}
\subsubsection{VSDE vs. CNF-only Dynamics}
We measure reconstruction accuracy on held-out bundles for each flow case and report the VSDE-augmented latent paths alongside the baseline CNF-only ODE. Across the Euler vortex, viscous shock tube, and incompressible cylinder experiments, the VSDE consistently improves RMSE and likelihood by steering trajectories back toward the learned manifold whenever the CNF starts to deviate.
\subsubsection{Integrator Comparison}
Inference stability is confirmed across Euler, Heun, RK4, and Dormand–Prince integrators for each flow, using overlay plots and energy statistics to quantify fidelity. After tuning the VSDE diffusion scale, all integrators reach similar accuracy, which confirms that the learned control dominates ensemble behavior while the choice of deterministic integrator only marginally affects fidelity in these flow regimes.

\section{Results}

This section presents experimental results corresponding to the hypotheses outlined in Section IV-A. We evaluate the proposed CNF–VSDE framework in terms of model expressiveness, the contribution of variational stochastic control, generalization across distinct flow regimes, and numerical stability under different ODE integration schemes.

Model performance is primarily quantified using mean absolute error (MAE) across all evaluated scenarios. MAE is selected due to its interpretability, as it measures average prediction error in the same physical units as the target variables, facilitating direct comparison with ground-truth fluid trajectories. In addition, MAE provides a linear error measure that does not disproportionately penalize large deviations, which is desirable in fluid dynamics applications where consistent accuracy is preferred over sensitivity to rare extreme events. Compared to squared-error metrics, MAE also offers increased robustness to occasional anomalies arising from turbulence or numerical noise.
\subsection{Expressiveness of ODE Backbones for Fluid Dynamics}
We first evaluate the performance of different ODE-based backbone models in capturing fluidic dynamics. Overall, the ODE-only approach is able to learn the dominant, mean flow behavior and produces smooth, physically consistent trajectories that align with the large-scale structure of the flow. In particular, the homogeneous CNF successfully converges to a stable solution that reflects the primary deterministic dynamics of the incompressible cylinder flow.

\begin{figure}[h!]
\centering
\includegraphics[width=1\linewidth]{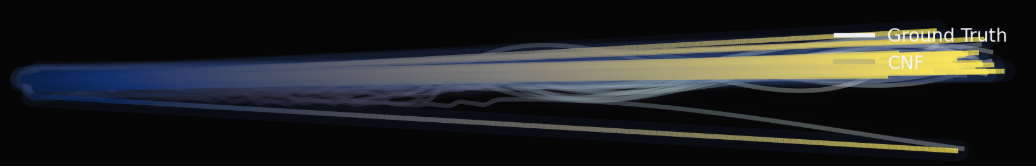}
\caption{Inference results of the ODE-only CNF model on the incompressible cylinder flow. Predicted trajectories are overlaid on the ground truth flow realizations. While the model accurately captures the overall mean flow structure and produces smooth, stable trajectories, the predictions collapse toward a deterministic solution. This behavior highlights the model’s inability to represent stochastic variations and multi-modal flow behavior, resulting in limited diversity compared to the ground truth dynamics.}
\label{fig:inc-cylinder-ode}
\end{figure}

However, this same determinism becomes a limitation when modeling stochastic fluid behavior. Although the model implicitly learns a probability distribution over trajectories, the absence of latent variables restricts its expressive capacity. As a result, the learned dynamics collapse toward a single dominant mode, making the model overly conservative and preventing it from capturing the full range of stochastic variations present in the system. This leads to an underrepresentation of flow variability and fine-scale randomness observed in the ground truth data. The inference results of the ODE-only model are shown in Figure~\ref{fig:inc-cylinder-ode}.

Additionally, we can compare the integrator performance using different numerical methods. We observed no significant difference in model accuracy or stability when using Euler, RK4, or Dormand-Prince integrators for the ODE backbone. This suggests that the learned dynamics are sufficiently smooth and well-behaved, allowing lower-order methods like Euler to perform comparably to higher-order schemes. The choice of integrator thus has minimal impact on the model’s ability to capture fluid behavior in this scenario. A comparison of integrator performance is shown in Appendix~\ref{apx:figures}, Figure~\ref{fig:integrator_comparison}.

\subsection{Impact of Variational Stochastic Control on Trajectory Fidelity}

To improve alignment with physical variability, we evaluate the effect of incorporating variational stochastic differential equation (VSDE) control into the CNF backbone. This analysis directly addresses whether stochastic augmentation enhances trajectory fidelity and uncertainty representation.

\begin{figure}[h!]
    \centering
    \includegraphics[width=1\linewidth]{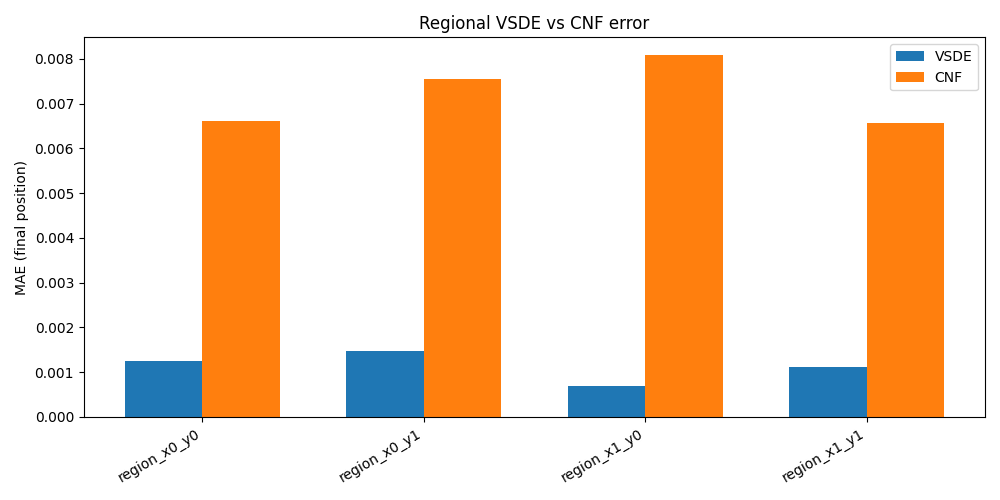}
    \caption{Regional comparison of mean absolute error (MAE) between the CNF-only and CNF–VSDE models, showing consistently lower error for VSDE across all spatial regions.}
    \label{fig:regional_error}
\end{figure}

Fig.~\ref{fig:regional_error} presents a spatial comparison of MAE between the CNF-only and CNF–VSDE models. Across all spatial regions, the VSDE-enhanced model consistently achieves lower reconstruction error. These improvements indicate that the learned posterior control actively compensates for deviations introduced by the frozen CNF drift, particularly in regions where stochastic effects play a significant role.

\begin{figure}[h!]
    \centering
    \includegraphics[width=1\linewidth]{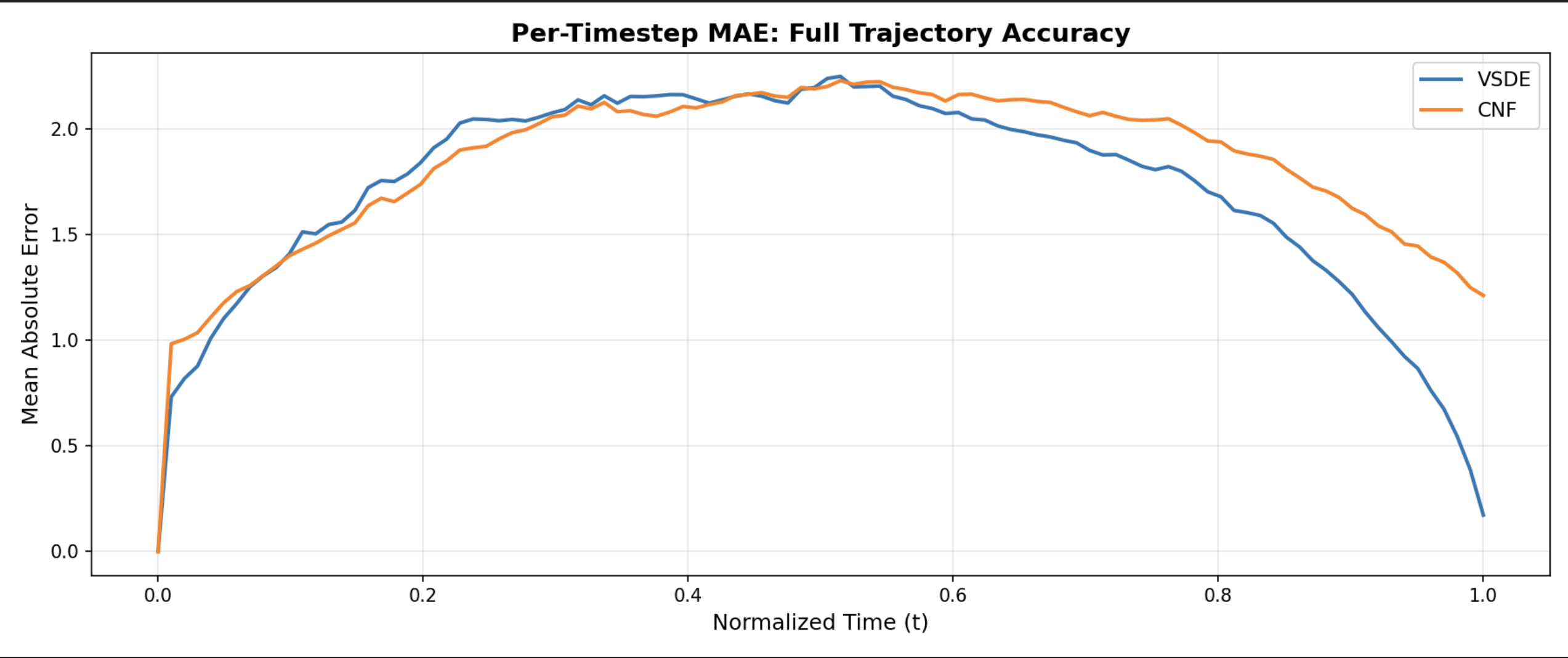}
    \caption{Comparison of ensemble trajectory behavior between the CNF-only and VSDE models. The CNF-only model produces relatively deterministic trajectories, whereas the VSDE framework generates ensembles exhibiting realistic dispersion. Although the VSDE model initially struggled to capture the accurate representation of turbulent and shear-dominated regions, the trajectories match the stochastic behavior observed in the CFD ground truth.}
    \label{fig:timestep_mae}
\end{figure}

\begin{figure}[h!]
    \centering
    \includegraphics[width=1\linewidth]{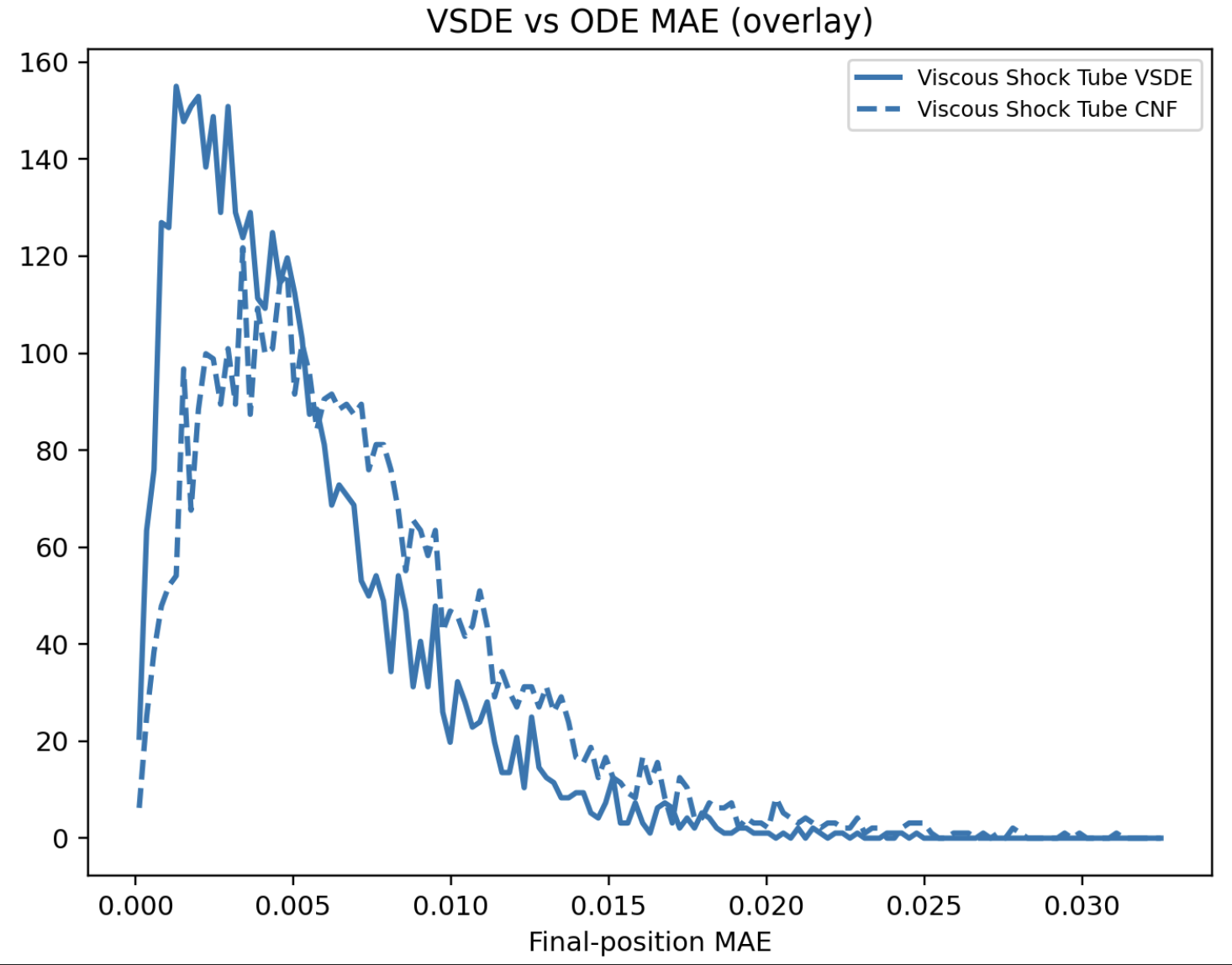}
    \caption{Distribution of final-position MAE for the viscous shock tube. VSDE control shifts the error distribution toward lower values relative to the CNF-only model, indicating improved trajectory fidelity without sacrificing ensemble diversity.}
    \label{fig:mae_error}
\end{figure}

\subsection{Physical Consistency in Velocity-Space Dynamics}

Physical consistency is further evaluated by comparing velocity phase-space structure and speed distributions against CFD reference statistics. As shown in Fig.~10, the ground-truth velocity field exhibits pronounced anisotropy in $(v_x, v_y)$ space and a broad, non-Gaussian speed distribution characteristic of shear- and turbulence-driven flow at early times. 

The CNF–VSDE model successfully reproduces these velocity-space patterns and preserves the empirical spread of particle speeds. In contrast, the CNF-only model collapses trajectories into a narrower region of velocity space, indicating an underrepresentation of physical variability. These results suggest that variational stochastic control enables the model to capture physically meaningful velocity distributions rather than merely approximating mean transport behavior.

A plot of generated trajectories overlays is included in Appendi~\ref{apx:figures}, Figure~\ref{fig:trajectory_overlay}.

\section{Discussion}
\subsection{Structural Roles of CNF and VSDE}
Our proposed framework decomposes the learning dynamic into two complementary components, with a deterministic, homogeneous backbone ODE implemented by the continuous normalizing flow (CNF), and a stochastic, inhomogeneous correction process implemented by the variational stochastic differential equation (VSDE). Our CNF implementation defines a single global, time-conditioned velocity field as
\[\frac{dz}{dt} = f_\theta (z,t,\alpha)\]

where the drift field, $f_{\theta}$ is globally parameterized and shared across all trajectories conditioned on the same flow parameters. The CNF implementation uses such a deterministic ODE to define a single, globally smooth flow map for a fixed conditioning context. Once the parameters and context are fixed, this ODE admits a unique solution for each initial condition, and all trajectories evolve under the same time-dependent vector field. As a result, the CNF is structurally biased toward capturing the mean advective behavior of the flow rather than trajectory-specific deviations.

This formulation is well-suited for learning globally consistent transport dynamics, but it inherently limits expressiveness in regimes where the true dynamics exhibit strong local variability. Because the governing equation is deterministic and homogeneous across trajectories, local mismatch between the learned vector field and the true dynamics accumulates monotonically along the flow. The diffeomorphic and stable nature of the CNF dynamics, while essential for accurate global transport, constrains the model’s ability to represent sharp transitions, localized mixing, or trajectory-dependent corrections. Introducing such effects would require redefining the global vector field, which would compromise the smoothness and consistency of the learned flow map. Consequently, the CNF lacks a mechanism to adapt locally without sacrificing its global structural properties.

The VSDE is explicitly designed to address this limitation by augmenting the CNF dynamics with trajectory-dependent and stochastic corrections. At each integration step, the frozen CNF drift $f_\theta(z,t)$ is evaluated and combined with a learned posterior control $u_\phi(z,t,\mathrm{ctx})$ to form the total drift. Brownian diffusion with a time-dependent scalar coefficient $g(t)$ is then injected during the update:
\begin{equation}
dz = \bigl[f_\theta(z,t) + u_\phi(z,t,\mathrm{ctx})\bigr]\,dt + \sqrt{2\,g(t)}\,dW_t.
\end{equation}
The control term $u_\phi$ breaks the homogeneity of the CNF dynamics by introducing trajectory-dependent forcing. For a fixed $(z,t)$, different posterior contexts yield different effective drifts. The diffusion term $g(t)$ introduces controlled variance into the dynamics, to perform stochastic exploration around the deterministic flow and mitigate error accumulation caused by incorrect drift alignment. These terms transform the governing equation from a single deterministic flow map into a family of stochastic trajectories conditioned on posterior information. Because the backbone drift $f_\theta$ is held fixed, the VSDE preserves the global flow structure learned by the CNF while locally perturbing the dynamics to correct trajectory-specific deviations.

These structural differences manifest clearly in the empirical results. Across all numerical integrators (see Figure~\ref{fig:integrator_comparison}), the CNF consistently exhibits higher final-position mean absolute error (MAE) than the VSDE. Although the neglible difference of the integrator confirms that our model learned a smooth, stable representation, rather then overfitting onto its training setting, the persistence of this gap across Euler, improved Euler, RK4, and DOPRI5 indicates that the dominant source of error is not numerical discretization, but the structure of the governing dynamics themselves. In particular, the deterministic and homogeneous nature of the CNF ODE leads to systematic error accumulation that cannot be mitigated by increased integration accuracy alone. While earlier figures illustrate consistent qualitative improvements, Table~\ref{tab:mae_summary} provides a quantitative summary of error reduction across representative flow regimes. In all cases, the VSDE achieves substantial reductions in final-position MAE relative to the CNF, with improvements exceeding 80\% across regimes and approaching 96\% in the shock-dominated Euler setting. Notably, these gains persist across both smooth and highly non-uniform flows, confirming that the observed improvements are not solver-dependent but arise from structural differences in the governing dynamics.

Spatially resolved error analysis further supports our discussion on this limitation. As shown in Fig.~\ref{fig:regional_error}, the CNF accumulates larger errors across all subregions, with the most pronounced discrepancies occurring in regions associated with sharp gradients and shock-driven dynamics. These regimes require localized, trajectory-specific adjustments that cannot be represented by a single global flow map. The VSDE significantly reduces error in these regions, which validates the posterior-conditioned control $u_\phi$ effectively compensates for localized dynamics that cannot be represented by a single homogeneous drift field.

\begin{table}[h!]
\centering
\small
\setlength{\tabcolsep}{4pt}
\caption{MAE comparison between CNF and VSDE.}
\label{tab:mae_summary}
\begin{tabular}{lccc}
\hline
\textbf{Flow Regime} & \textbf{CNF MAE} & \textbf{VSDE MAE} & \textbf{Red. (\%)} \\
\hline
Viscous Shock        & 0.0711 & 0.0119 & 83.2 \\
Euler Vortex         & 1.5828 & 0.0659 & 95.8 \\
Incompressible Cyl.  & 1.7010 & 0.2646 & 84.4 \\
\hline
\end{tabular}
\end{table}

This behavior is also evident in the distributional analysis of final-position error. The CNF-only model exhibits a broad, heavy-tailed MAE distribution, indicating systematic drift and accumulated error along trajectories. When overlaid with the VSDE results, the VSDE distribution is noticeably more concentrated near zero error, reflecting improved trajectory alignment. The difference distribution (Fig.~\ref{fig:mae_error}) is strongly centered on positive values, indicating that the CNF incurs higher final-position error for the majority of trajectories. The pronounced peak near a small positive offset reflects a systematic improvement introduced by the VSDE rather than sporadic or outlier-driven gains. The relatively light tails further indicate that the VSDE does not introduce significant new failure modes or instability relative to the CNF.

\subsection{Error Characteristics and Sources}

The dominant error patterns observed in the results stem from inherent limitations of the governing differential equations rather than numerical discretization. In the CNF formulation, small local discrepancies between the learned drift and the true dynamics propagate coherently along deterministic trajectories, leading to systematic error growth. These effects are most pronounced in regions with shocks, diffusion-dominated behavior, and sharp spatial gradients, where the true dynamics exhibit localized variability that cannot be captured by a globally smooth flow map. The persistence of this error across numerical integration schemes—including higher-order solvers—further confirms that the discrepancies arise from structural bias in the ODE formulation rather than from insufficient numerical accuracy.

The VSDE reduces these errors by augmenting the deterministic backbone with an additive stochastic correction, but residual error remains due to intrinsic limits of this formulation. Because the posterior control enters additively, the VSDE can only locally perturb the deterministic flow rather than redefine it. As a result, large structural mismatches in the frozen backbone drift cannot be fully eliminated, resulting in an irreducible error floor. The stochastic component introduces a bias–variance tradeoff inherent to SDE dynamics. While diffusion enables flexibility and prevents trajectories from remaining trapped along incorrect deterministic paths, it also introduces variance that limits achievable precision. Although the time-decaying diffusion schedule suppresses endpoint noise, early-time stochastic perturbations can propagate forward through the dynamics.

\section{Conclusion}

\subsection{Summary of Contributions}
The project aims to model physical representations with efficient generative models, making it suitable and effective for applications in fluid modeling. We examined the structural limitations of continuous normalizing flows when used to model complex, gradient varying dynamics and demonstrated how a variational stochastic differential equation formulation can address these limitations. By interpreting CNF and VSDE through their governing differential equations, we showed that deterministic, homogeneous flow models are inherently prone to systematic error accumulation, while trajectory-conditioned stochastic corrections enable localized adaptation without sacrificing global coherence. Empirical results across multiple integration schemes and spatial regimes support this interpretation, with the VSDE consistently reducing error relative to the CNF. The stochastic component introduces a bias–variance tradeoff inherent to SDE dynamics. The impacts of model structure in learned dynamical systems are identified as critical aspects in the behaviour and quality of results and motivate further exploration of stochastic and controlled extensions to deterministic flow-based models.

\subsection{Future Work}
While the VSDE effectively corrects trajectory-level deviations by additively augmenting a frozen deterministic drift, this formulation inherently limits expressiveness in regimes where the backbone dynamics themselves are structurally misrepresented. In such cases, additive control may be insufficient to capture state-dependent changes in the underlying flow geometry and our future exploration could build upon the multiplicative or state-dependent control mechanisms that more fundamentally reshape the governing dynamics. The use of a time-decaying diffusion schedule was not informed with a realistic theoretically optimal policy. Further work could investigate adaptive or learned diffusion schedules that dynamically balance stochastic exploration and precision across time, potentially improving accuracy while retaining stability.

\newpage

\appendices
\onecolumn

\section{Selected Additional Figures}\label{apx:figures}
\begin{figure}[h!]
    \centering
    \includegraphics[width=0.5\linewidth]{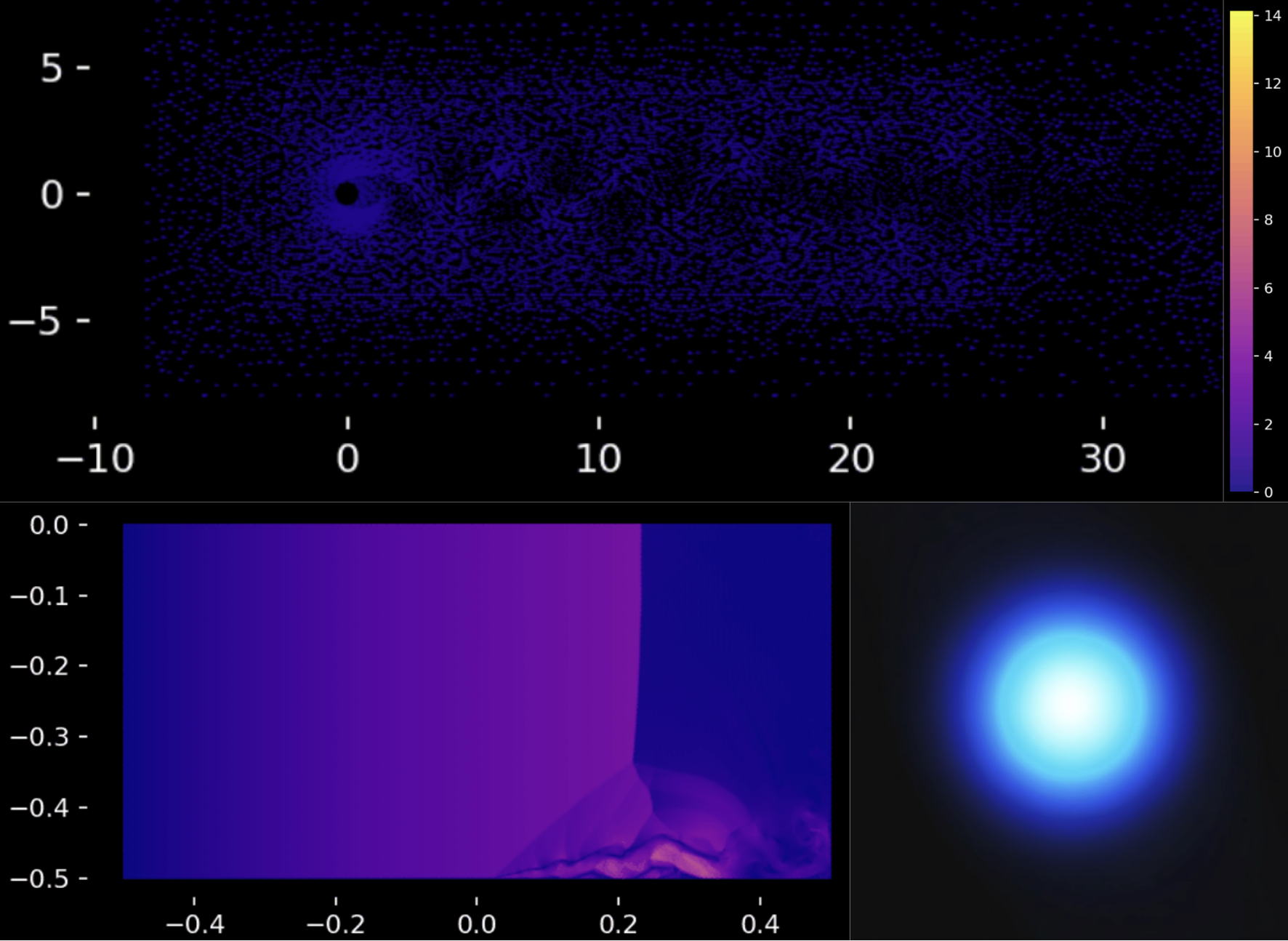}
    \caption{\textbf{Flow Profiles for Trajectory Generation.} (a) and (b) are velocity profiles on the top  and bottom right respectively. (c) is located bottom right, a pressure profile visualized via Paraview \cite{paraview}.  (a) \emph{Shock Tube} configuration showing discontinuity complex behavior near the wall. (b) \emph{Cylinder Flow}, where the flow impinges on a cylinder, producing a radial velocity distribution and oscillatory trail downstream. (c) \emph{Euler Vortex}, illustrating rotational shear flow with centerline velocity $U_\text{max}$. }
    \label{fig:flow_vis}
\end{figure}

\begin{figure}[h!]
    \centering
    \includegraphics[width=0.5\linewidth]{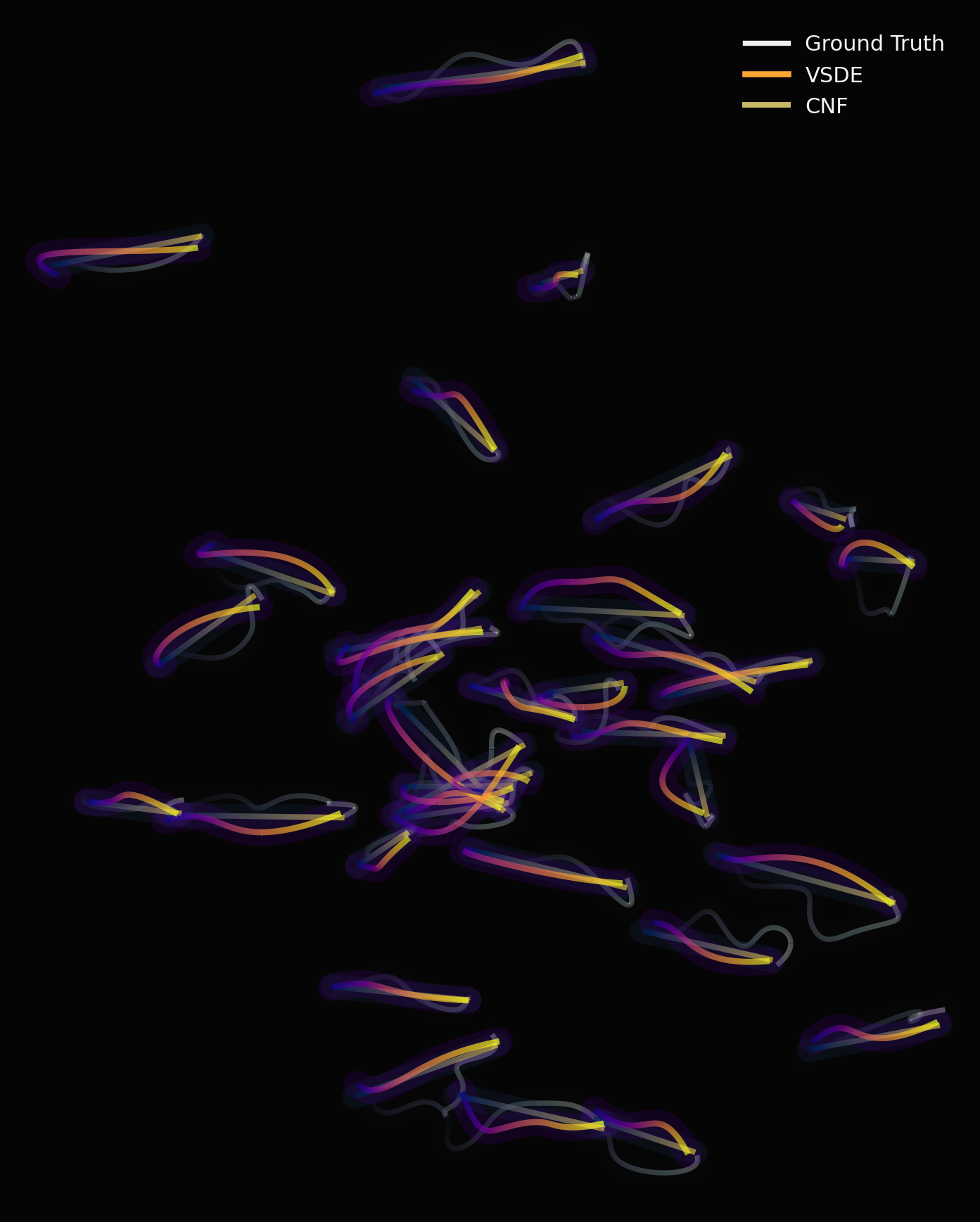}
    \caption{Qualitative comparison of trajectory reconstructions for the viscous shock tube.
Ground-truth particle trajectories (gray) are overlaid with predictions from the CNF–VSDE model (purple) and the CNF-only model (yellow). The VSDE trajectories more closely follow the geometry and dispersion of the ground truth, particularly in regions of strong curvature and shear, while the CNF-only model exhibits accumulated drift and reduced variability.}
    \label{fig:trajectory_overlay}
\end{figure}

\begin{figure}[h!]
    \centering
    \includegraphics[width=0.5\linewidth]{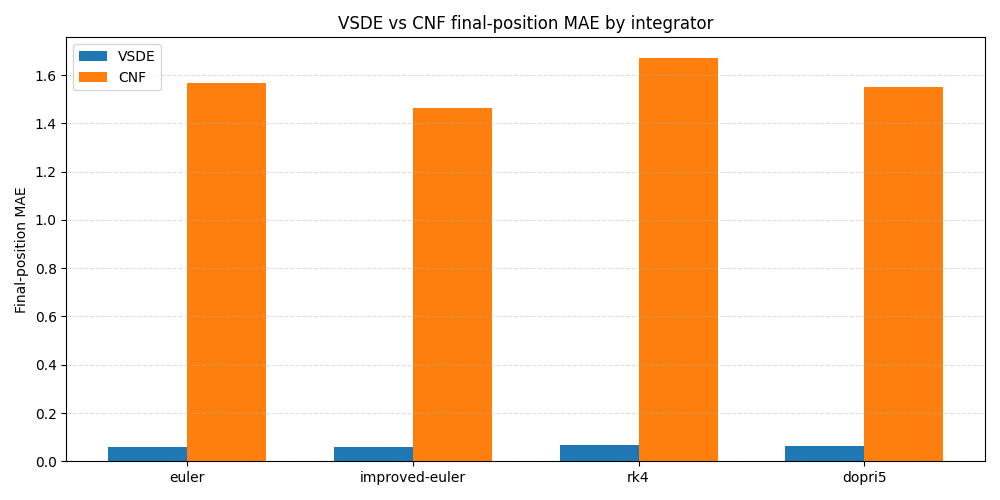}
    \caption{Comparison of mean absolute error (MAE) across different ODE integrators (Euler, Heun, RK4, Dormand-Prince) for the ODE-only CNF model on the incompressible cylinder flow. The results indicate negligible differences in accuracy and stability among the integrators, suggesting that the learned dynamics are sufficiently smooth for lower-order methods to perform comparably to higher-order schemes.}
        
    \label{fig:integrator_comparison}
\end{figure}

\newpage
\section{PyFR Orchestration} \label{apx:pyfr}

Simulations were orchestrated within a Jupyter Notebook environment to facilitate reproducible execution, automated workflow control, GPU acceleration via CUDA, and seamless integration with downstream post-processing and dataset management procedures.

The output of the simulation pipeline is a set of ground-truth physical trajectories representing the evolution of flow variables over time. Each simulation is initialized from a mesh file (\texttt{.msh}) defining the domain discretization and a solver configuration file (\texttt{.ini}) specifying physical parameters and temporal integration settings; these inputs are converted into \texttt{PyFR}'s internal formats prior to execution. Post-processing converts the solver state outputs into structured numerical arrays that preserve spatial and temporal coherency, yielding discrete samples of an underlying continuous-time dynamical system governed by Eq.~\eqref{eq:Navier-Stokes}. These datasets provide the reference trajectories used to train and validate the learned dynamical models~\cite{witherden2025pyfr}.

To support reproducibility, all data, simulation scripts, and configuration files are provided alongside the accompanying GitHub repository.
\newpage
\section{Implementation and Reproducibility} \label{apx:implementation}
All code lives in \texttt{\url{github.com/haysonc/pivonet}}. Clone the repository, install dependencies, and run the VSDE model with the following steps:
\begin{enumerate}
\item Set up the virtual environment and install the requirements:
\begin{verbatim}
python -m venv venv
source venv/bin/activate
pip install -r requirements.txt
\end{verbatim}
\item Use the \texttt{pivo} CLI to orchestrate training/inference or call the flow module:\\
\begin{verbatim}
pivo
\end{verbatim}
\end{enumerate}
Dataset from \texttt{PyFR} is uploaded on Google Drive: \texttt{\url{https://drive.google.com/drive/folders/13ykGleAmZTNAz1lhR0x6FekYqyGYbper?usp=sharing}}. You can download and extract it into the \texttt{data/} folder in the repository root.

A sample repository structure is as follows:
\begin{verbatim}
pivonet/
|-- data/
|   |-- 2d-euler-vortex/
|   `-- ...
|-- src/
|   |-- models/
|   |-- utils/
|   `-- ...
`-- ...
\end{verbatim}
For more detailed instructions, refer to the \texttt{README.md} in the repository root.
The code style is enforced with \texttt{ruff} and is configured via \texttt{pyproject.toml} ; all configuration values follow YAML-like semantics with leading keys for clarity, so keep any new scripts aligned with those defaults.

\newpage
\section{Hyperparameter Summary} \label{apx:hyperparams}

The tables below summarize the representative hyperparameters for each flow case we evaluate, including trajectory sampling, CNF backbone training, VSDE controller fitting, and inference diagnostics.  

We first provide an overview table describing the meaning of each hyperparameter.

\begin{table}[h!]
    \centering
    \caption{Hyperparameter descriptions for all flows.}
    \begin{tabular}{ll}
        \hline
        \textbf{Component} & \textbf{Description} \\
        \hline
        \multicolumn{2}{l}{\textbf{Trajectory}} \\
        Particles & Number of simulated particles in the flow \\
        Steps & Number of integration steps for trajectory sampling \\
        \(\Delta t\) & Time step for integration \\
        \hline
        \multicolumn{2}{l}{\textbf{CNF}} \\
        Batch size & Number of samples per training batch \\
        Epochs & Number of training passes over the dataset \\
        Learning rate & Step size for optimizer updates \\
        Hidden units & Number of neurons per hidden layer \\
        Depth & Number of hidden layers \\
        Context dim & Dimensionality of context vector for conditioning \\
        Limit & Maximum number of training samples per epoch \\
        \hline
        \multicolumn{2}{l}{\textbf{VSDE}} \\
        Batch size & Number of samples per training batch \\
        Epochs & Number of training passes over the dataset \\
        Learning rate & Step size for optimizer updates \\
        Particles & Number of controlled particles for VSDE fitting \\
        Steps & Number of integration steps for VSDE dynamics \\
        Control cost & Weight for control penalty in the loss \\
        Learnable diffusion & Whether diffusion coefficient is learned \\
        \hline
        \multicolumn{2}{l}{\textbf{Inference}} \\
        Particles & Number of ensemble particles for inference \\
        Steps & Number of integration steps during inference \\
        Diagnostics & Types of outputs generated for evaluation \\
        \hline
    \end{tabular}
\end{table}

\subsection{Euler Vortex}
\begin{table}[h!]
    \centering
    \caption{Hyperparameters for the Euler Vortex flow.}
    \label{tab:hyperparams_euler}
    \begin{tabular}{ll}
        \hline
        \textbf{Component} & \textbf{Value} \\
        \hline
        \multicolumn{2}{l}{\textbf{Trajectory}} \\
        Particles & 4096 \\
        Steps & 240 \\
        \(\Delta t\) & 0.01 \\
        \hline
        \multicolumn{2}{l}{\textbf{CNF}} \\
        Batch size & 512 \\
        Epochs & 8 \\
        Learning rate & 0.002 \\
        Hidden units & 64 \\
        Depth & 3 \\
        Context dim & 3 \\
        Limit & 1024 \\
        \hline
        \multicolumn{2}{l}{\textbf{VSDE}} \\
        Batch size & 2048 \\
        Epochs & 50 \\
        Learning rate & 0.01 \\
        Particles & 64 \\
        Steps & 120 \\
        Control cost & 1.0 \\
        Learnable diffusion & Yes \\
        \hline
        \multicolumn{2}{l}{\textbf{Inference}} \\
        Particles & 64 \\
        Steps & 120 \\
        Diagnostics & Overlay outputs + phase plots \\
        \hline
    \end{tabular}
\end{table}

\subsection{Viscous Shock Tube}
\begin{table}[h!]
    \centering
    \caption{Hyperparameters for the Viscous Shock Tube flow.}
    \label{tab:hyperparams_vis}
    \begin{tabular}{ll}
        \hline
        \textbf{Component} & \textbf{Value} \\
        \hline
        \multicolumn{2}{l}{\textbf{Trajectory}} \\
        Particles & 1024 \\
        Steps & 240 \\
        \(\Delta t\) & 0.01 \\
        \hline
        \multicolumn{2}{l}{\textbf{CNF}} \\
        Batch size & 512 \\
        Epochs & 8 \\
        Learning rate & 0.001 \\
        Hidden units & 64 \\
        Depth & 3 \\
        Context dim & 3 \\
        Limit & 512 \\
        \hline
        \multicolumn{2}{l}{\textbf{VSDE}} \\
        Batch size & 128 \\
        Epochs & 4 \\
        Learning rate & 0.005 \\
        Particles & 4 \\
        Steps & 60 \\
        Control cost & 1.0 \\
        Learnable diffusion & Yes \\
        \hline
        \multicolumn{2}{l}{\textbf{Inference}} \\
        Particles & 32 \\
        Steps & 60 \\
        Diagnostics & Overlay outputs + phase plots \\
        \hline
    \end{tabular}
\end{table}
\newpage
\subsection{Incompressible Cylinder}
\begin{table}[h]
    \centering
    \caption{Hyperparameters for the Incompressible Cylinder flow.}
    \label{tab:hyperparams_inc}
    \begin{tabular}{ll}
        \hline
        \textbf{Component} & \textbf{Value} \\
        \hline
        \multicolumn{2}{l}{\textbf{Trajectory}} \\
        Particles & 1024 \\
        Steps & 240 \\
        \(\Delta t\) & 0.01 \\
        \hline
        \multicolumn{2}{l}{\textbf{CNF}} \\
        Batch size & 512 \\
        Epochs & 8 \\
        Learning rate & 0.001 \\
        Hidden units & 64 \\
        Depth & 3 \\
        Context dim & 3 \\
        Limit & 512 \\
        \hline
        \multicolumn{2}{l}{\textbf{VSDE}} \\
        Batch size & 128 \\
        Epochs & 4 \\
        Learning rate & 0.005 \\
        Particles & 4 \\
        Steps & 60 \\
        Control cost & 1.0 \\
        Learnable diffusion & Yes \\
        \hline
        \multicolumn{2}{l}{\textbf{Inference}} \\
        Particles & 32 \\
        Steps & 60 \\
        Diagnostics & Overlay outputs + phase plots \\
        \hline
    \end{tabular}
\end{table}

The hyperparameters were selected to balance training stability and inference accuracy across different flow regimes. CNF and VSDE settings scale with flow complexity, while inference configurations ensure sufficient ensemble diversity for performance evaluation. All values are also available in the repository under \texttt{src/experiments/} in the \texttt{.yaml} configuration files. Although the version uploaded to GitHub may have parameters reduced to improve runtime for the demo.

\newpage


\begin{thebibliography}{10}

\bibitem{Squires2005}
T.~M. Squires and S.~R. Quake, ``Microfluidics: Fluid physics at the nanoliter scale,'' {\em Reviews of Modern Physics}, vol.~77, pp.~977--1026, 2005.

\bibitem{Ermak1978}
D.~L. Ermak and J.~A. McCammon, ``Brownian dynamics with hydrodynamic interactions,'' {\em Journal of Chemical Physics}, vol.~69, pp.~1352--1360, 1978.

\bibitem{ODETextbook}
R.~L. Devaney and S.~G. Krantz, {\em Differential Equations: An Introduction to Modern Methods}.
\newblock Brooks/Cole, 2nd~ed., 2007.

\bibitem{Taylor1953}
G.~I. Taylor, ``Dispersion of soluble matter in solvent flowing slowly through a tube,'' {\em Proceedings of the Royal Society of London. Series A}, vol.~219, pp.~186--203, 1953.

\bibitem{Chen2018}
R.~T.~Q. Chen, Y.~Rubanova, J.~Bettencourt, and D.~Duvenaud, ``Neural ordinary differential equations,'' in {\em Advances in Neural Information Processing Systems (NeurIPS)}, 2018.

\bibitem{Grathwohl2019}
W.~Grathwohl, R.~T.~Q. Chen, J.~Bettencourt, I.~Sutskever, and D.~Duvenaud, ``Ffjord: Free-form continuous dynamics for scalable reversible generative models,'' in {\em Proc. International Conference on Machine Learning (ICML)}, 2019.

\bibitem{Dinh2014}
L.~Dinh, D.~Krueger, and Y.~Bengio, ``Nice: Non-linear independent components estimation,'' in {\em ICLR Workshop Track}, 2014.

\bibitem{Dinh2017}
L.~Dinh, J.~Sohl-Dickstein, and Y.~Bengio, ``Density estimation using real nvp,'' in {\em Proc. International Conference on Learning Representations (ICLR)}, 2017.

\bibitem{Kingma2018}
D.~P. Kingma and P.~Dhariwal, ``Glow: Generative flow with invertible 1x1 convolutions,'' in {\em Advances in Neural Information Processing Systems (NeurIPS)}, 2018.

\bibitem{Rezende2015}
D.~J. Rezende and S.~Mohamed, ``Variational inference with normalizing flows,'' in {\em Proc. 32nd International Conference on Machine Learning (ICML)}, 2015.

\bibitem{Goodfellow2016}
I.~Goodfellow, Y.~Bengio, and A.~Courville, {\em Deep Learning}.
\newblock MIT Press, 2016.

\bibitem{adam}
D.~P. Kingma and J.~Ba, ``Adam: A method for stochastic optimization,'' {\em arXiv preprint arXiv:1412.6980}, 2015.

\bibitem{Cho2014}
K.~Cho, B.~van Merrienboer, C.~Gulcehre, D.~Bahdanau, F.~Bougares, H.~Schwenk, and Y.~Bengio, ``Learning phrase representations using rnn encoder--decoder for statistical machine translation,'' in {\em EMNLP}, 2014.

\bibitem{witherden2025pyfr}
F.~D. Witherden, B.~C. Vermeire, P.~E. Vincent, {\em et~al.}, ``{PyFR v2.0.3: Towards industrial adoption of scale-resolving simulations},'' {\em Computer Physics Communications}, vol.~311, p.~109567, 2025.

\bibitem{paraview}
{Ahrens, James and Geveci, Berk and Law, Charles}, ``Paraview.'' \url{https://www.paraview.org/}, 2005.
\newblock Version 5.11.

\end{thebibliography}
\end{document}